# Two-sided receptivity to conversational AI agents in online dating: Bilingual survey data from Fledge.Love

Daria Leshchikova[1] [0009-0003-2916-4244], Valentina V. Kuskova[2]* [0000-0003-4716-2544], Dmitry Zaytsev[2] [0000-0002-0902-3896], Valerii Klimov[1] [0009-0007-0373-9854]

[1] Fleamily, Inc., Delaware, USA

[2] Lucy Family Institute for Data & Society, University of Notre Dame, Notre Dame, Indiana, USA.

*Corresponding author: vkuskova@nd.edu

## Abstract

Autonomous conversational agents and generative-AI features are being added to online dating platforms faster than public evidence about user attitudes can accumulate, and the scarcest evidence concerns the receiving side: how people react when the profiles, messages, or conversation partners they encounter are machine-generated. We release two anonymized survey datasets collected from active users of Fledge.Love, a dating platform serving an international user base. The first (N = 2,617; Russian and English forms) measures receptivity to autonomous conversational agents with a seven-item battery that separates the principal role (deploying one's own agent) from the counterpart role (encountering someone else's), plus six ordinal covariates and two auxiliary items. The second (N = 2,894) measures interest in three passive generative-AI features. The release includes model-derived scores for 2,499 complete cases, a bilingual codebook, a documented anonymization pipeline with a k-anonymity audit, executable analysis notebooks, and canonical outputs, supporting reuse in human-AI communication, recommender-systems, and cross-cultural technology-acceptance research.

## Background & Summary

Matching platforms are an early commercial frontier for interpersonal AI: profile summarization, message suggestions, and fully autonomous conversational agents are moving from prototypes into products [1-2]. Research access to user-attitude data has not kept pace. Existing public instruments measure general attitudes toward artificial intelligence [3-5] or acceptance of technology in the abstract [6-7]; what they do not capture is the structure specific to communication delegation [8], where every user occupies two roles at once. A person may deploy an agent to converse on their behalf, and the same person may encounter agents deployed by others. Attitudes toward these two roles need not coincide, and design decisions - disclosure rules, opt-in mechanics, routing - depend on both. To our knowledge, no public microdata measure the two roles within the same respondents, in any language, and survey data from dating-platform user bases of any kind are rare because access normally requires commercial agreements.

This release makes such data public. Two survey instruments were fielded by Fledge.Love (https://fledge.love), a dating platform with an international user base, to its active users via an in-app prompt. Instrument B (N = 2,617; fielded November-December 2025 in parallel Russian and English forms) measures receptivity to autonomous conversational agents through seven ordinal items constructed around the two roles: four principal-role items (first reaction to the concept, attitude to configuring the agent, willingness to deploy one's own agent, and overall product value) and three counterpart-role items (willingness to engage another person's agent, reaction to agent-to-agent pre-conversation, and reaction to mixed human-agent group chat). Six ordinal covariates capture gender, age band, platform tenure, match volume, match-to-conversation experience, and affect about stalled conversations, alongside two auxiliary categorical items on why conversations stall and how respondents cope. Instrument A (N = 2,894; fielded March-April 2026) measures interest in three passive generative-AI features - an AI profile summary, AI conversation tips, and an AI couple description - together with a best-of-three choice, a browsing-interest item, and demographics including self-reported AI usage. The samples are unlinked.

The released files are the analysis dataset of record. Anonymization consisted of dropping contact information unread, removing all free text, coarsening timestamps to ISO week, auditing k-anonymity at k >= 5 with one documented local recoding, and shuffling row order under opaque identifiers. It was applied before any derived quantity was computed, so every released derived value reproduces exactly from the released data via the released code. The deposit, therefore, contains, alongside the three data files, an anonymization audit log; a codebook documenting both language forms verbatim with every coding map; a model-derived scores file (expected a posteriori factor scores and endorsement propensities under a two-dimensional graded response model) for the 2,499 records complete on all items and covariates; three executable notebooks that regenerate every derived quantity; the anonymization pipeline itself; and the canonical outputs those notebooks produce, with SHA-256 checksums throughout.

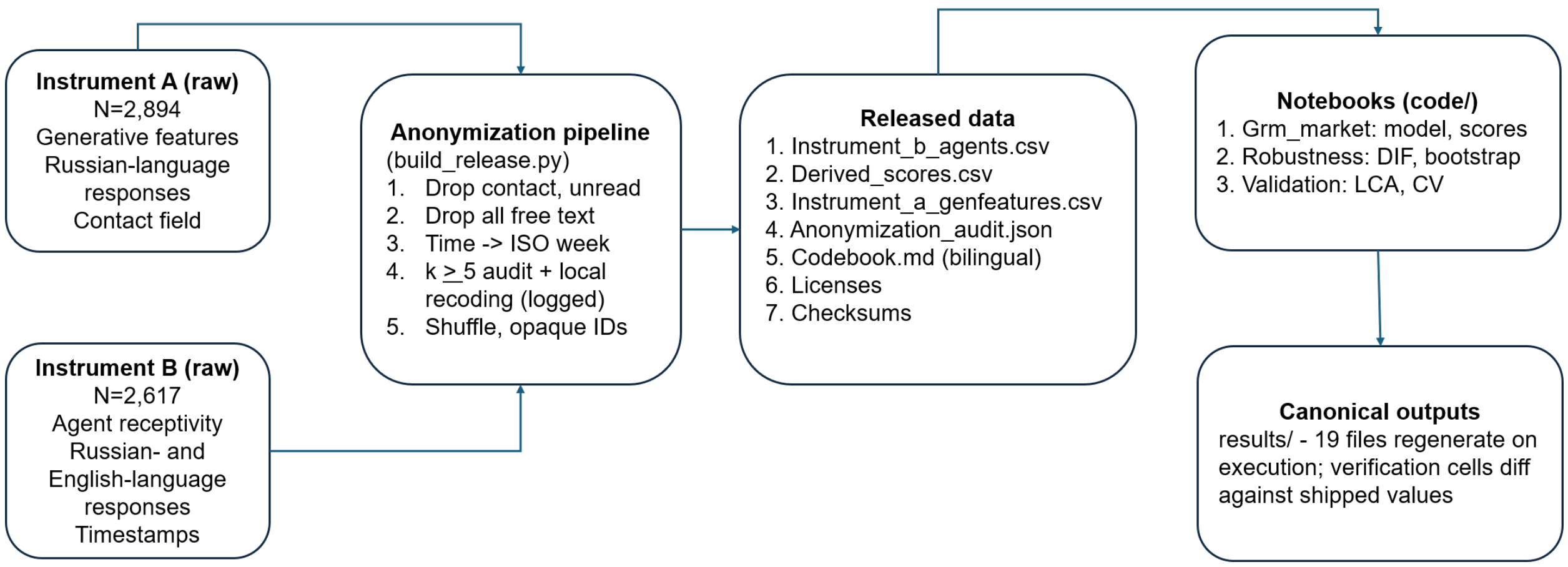

***Figure 1: Structure of the release. Two raw survey exports pass through the anonymization pipeline into the released data files; the released notebooks read only the released files and regenerate every canonical output, each ending with a verification cell that compares its outputs to the shipped values.***

We anticipate reuse in several communities. Human-AI communication researchers gain, to our knowledge, the first public item battery separating send-side from receive-side receptivity, with documented measurement properties. Recommender-systems researchers gain per-user receive-side receptivity quantities suitable as modeling targets or benchmarks for agentic and reciprocal recommendation [9]. Cross-cultural technology-acceptance researchers gain a bilingual instrument with item-level differential-functioning results documenting exactly where the two language forms are and are not comparable. And because the dataset is small, complete, and ships with a full estimation pipeline and known outputs, it is suitable as teaching material for item response theory, measurement invariance, and latent class analysis.

# Methods

## Platform and recruitment

Both surveys were designed, fielded, and collected by Fledge.Love as voluntary, self-administered online questionnaires. Recruitment used an in-app prompt shown to active users; no incentive was offered and declining to participate carried no consequence. Instrument B ran from November 12 to December 15, 2025, with parallel Russian and English forms. Respondents received the form matching their app language, and the released language field records which form each respondent completed. Instrument A ran from March 22 to April 1, 2026, in Russian. Submission timing is released at ISO-week resolution.

## Instrument B: agent receptivity

The seven receptivity items are reproduced verbatim in both language forms in Tables 1–2, together with the analysis codes distributed in the released files. Items Y1–Y3 and Y7 place the respondent in the principal role (configuring or deploying an agent of their own); items Y4–Y6 place the respondent in the counterpart role (encountering agents deployed by others). Y7 is a 1–10 product-value rating binned to five levels; all other items are closed-choice.

**Table 1. Instrument B receptivity items Y1–Y4: verbatim response options, both language forms, with released analysis codes.**

| Item | Stem (paraphrase) | English option (verbatim) | C | Russian option (verbatim) | C |
|---|---|---|---|---|---|
| Y1 | First reaction to the agent concept | Very interesting | 2 | Очень интересно | 2 |
| | | Neutral | 1 | Нейтрально | 1 |
| | | Not interesting | 0 | Не интересно | 0 |
| | | It didn't seem trustworthy | 0 | Не вызвало доверия | 0 |
| Y2 | Attitude to configuring the agent’s activity and tone | Useful: it will save me time | 2 | Полезно: сэкономит мне время | 2 |
| | | Too complicated: I don't feel like figuring it out | 1 | Сложно: лень в этом разбираться | 1 |
| | | Risky: what if it writes something inappropriate | 0 | Опасно: вдруг напишет лишнего | 0 |
| Y3 | Deploying one’s own agent (“the agent talks with a potential partner while you are away”) | I want to try it | 2 | Хочу попробовать | 2 |
| | | I might try it | 1 | Возможно, попробую | 1 |
| | | I don't want an agent responding for me | 0 | Не хочу чтобы за меня отвечал агент | 0 |
| Y4 | Engaging another person’s agent (“someone’s AI agent responded to you”) | I would engage in a conversation with the agent | 2 | Вовлекусь в диалог с агентом | 2 |
| | | I might engage | 1 | Возможно, вовлекусь | 1 |
| | | I would react negatively | 0 | Отнесусь негативно | 0 |
| | | That's weird, I only want to chat with a real person | 0 | Странно, хочу общаться только с человеком | 0 |

**Table 2. Instrument B items Y5–Y6 (verbatim options) and Y7 (rating bins).**

| Item | Stem (paraphrase) | English option (verbatim) | C | Russian option (verbatim) | C |
|---|---|---|---|---|---|
| Y5 | Agent-to-agent pre-conversation before the humans join | Super idea | 3 | Супер идея | 3 |
| | | Funny, I would watch their conversation | 2 | Забавно, посмотрел бы на их разговор | 2 |
| | | Doubtful, why are we needed then? | 1 | Сомнительно, зачем тогда мы? | 1 |
| | | Strange and frightening | 0 | Странно и пугающе | 0 |
| Y6 | Mixed human–agent group chat | Super idea | 2 | Супер идея | 2 |
| | | Funny, I would participate | 1 | Забавно, я бы поучаствовал(а) | 1 |
| | | Doubtful, why are we needed then? | 0 | Сомнительно, зачем тогда мы? | 0 |
| | | — | | Странно и пугающе | 0 |

Note. The Russian form of Y6 offers a fourth option (“Странно и пугающе”, coded 0) not present in the English form; both forms of Y1 and Y4 map two rejection-flavored options to code 0. The released codebook (CODEBOOK.md) and the released pipeline (build_release.py) are the machine-readable sources of these mappings.

Y7: respondents rate overall product value on a 1–10 scale; released code bins ratings as 1–2→0, 3–4→1, 5–6→2, 7–8→3, 9–10→4. The raw 1–10 value is also released (Y7_raw_1to10).

## Covariates and auxiliary items (Instrument B)

Six ordinal covariates follow the items; Table 3 provides every response option verbatim in both forms with the released codes. Two auxiliary closed-choice items ask why conversations stall and what the respondent does about stalled matches; their options (Table 4) are released verbatim in the language of the form, uncoded, as decay_reason and coping.

**Table 3. Instrument B covariates: verbatim response options, both forms, with released codes.**

| Variable | Construct | English option (verbatim) | C | Russian option (verbatim) | C |
|---|---|---|---|---|---|
| female | Gender | Male | 0 | Мужской | 0 |
| | | Female | 1 | Женский | 1 |
| age | Age band | 18-24 | 0 | 18-24 | 0 |
| | | 25-34 | 1 | 25-34 | 1 |
| | | 35-44 | 2 | 35-44 | 2 |

| | | 45+ | 3 | 45+ | 3 |
|---|---|---|---|---|---|
| tenure | Time on the platform | It's my first month | 0 | Первый месяц | 0 |
| | | A few months | 1 | Несколько месяцев | 1 |
| | | Over a year | 2 | Больше года | 2 |
| match_enough | Perceived match volume | I have very few matches | 0 | У меня мало мэтчей | 0 |
| | | Not very many | 1 | Не очень много | 1 |
| | | Enough | 2 | Достаточно | 2 |
| | | Too many | 3 | Слишком много | 3 |
| conv | Share of matches becoming conversations | Very few | 0 | Очень мало | 0 |
| | | Less than half | 1 | Меньше половины | 1 |
| | | More than half | 2 | Больше половины | 2 |
| | | Almost all of them | 3 | Почти все | 3 |
| neg_emo | Negative feelings about stalled matches | They don't cause any negative feelings | 0 | Не вызывают негатива | 0 |
| | | I feel neutral | 1 | Отношусь нейтрально | 1 |
| | | Yes, they do | 2 | Да, вызывают | 2 |

**Table 4. Auxiliary categorical items: verbatim options by form (released uncoded; ordering by frequency).**

| Variable | Construct | English options (verbatim) | Russian options (verbatim) |
|---|---|---|---|
| decay_reason | Attributed reason conversations stall | They don't reply to me | Жду первый шаг от партнера |
| | | I'm waiting for the partner to make the first move | Мне не отвечают |
| | | I lack the energy/time | Нет ресурса |
| | | I don't know how to start | Не знаю как начать |
| | | Too many active chats | Слишком много диалогов |
| | | It doesn't happen to me | У меня так не бывает |
| coping | Strategy for stalled matches | Try to engage the person in conversation | Пусть висят без цели |
| | | Let them remain inactive | Оставлю на будущее |
| | | Delete the match (Unmatch) | Удалить мэтч |
| | | I'll save them for later | Попробовать разговорить человека |

## Instrument A: generative-feature interest

Instrument A presented three passive generative-feature concepts: an AI summary of the respondent's profile tests (idea1_profile), AI conversation tips (idea2_tips), and an AI-written description of a prospective couple (idea3_couple), each rated on a three-level interest scale (not interesting 0; interesting 1; very interesting 2). A free-text "other" response was possible for each rating; where used, the released rating is missing and the corresponding *_nonresponse indicator is 1 (the text is not released). A forced-choice item selects the best of the three concepts (best_of_three: idea1/idea2/idea3/none), and a four-level item records whether such features would increase browsing interest (interest_increase). Demographics comprise gender, age band, app usage frequency (usage_freq), and self-reported AI-service usage (ai_usage); codings appear in the released codebook and in Table 7.

## Anonymization

Five operations were applied to produce the released files, in order: (1) the contact field collected by Instrument A was dropped without being read; (2) all free-text content was removed, with free-text "other" ratings in Instrument A represented by a nonresponse indicator; (3) submission timestamps were coarsened to ISO week; (4) a k-anonymity audit ($k \geq 5$) was run on gender × age band × language; (5) rows were shuffled with a fixed seed and assigned opaque respondent identifiers. The audit found a single violating cell (English-form women aged 45+, $n = 3$), resolved by local recoding: the two oldest age bands were merged to "35+" for English-form women (12 records' band label; 3 records' numeric age code). The full audit log is released as anonymization_audit.json [Data Citation 1]. The pipeline script (build_release.py) is released and re-derives the analysis coding from the raw exports, asserting row-level equality before writing any file; the released files are the analysis dataset of record — the recoding was applied before all derived quantities were computed.

### Derived scores

The file derived_scores.csv provides, for each of the 2,499 records complete on all seven items and six covariates, model-derived quantities computed as a processing step on the released data [Data Citation 1]: expected a posteriori (EAP) scores on two latent dimensions (theta_send from the principal-role items, theta_recv from the counterpart-role items) and four endorsement propensities: the model-implied probabilities of endorsing deployment (from item Y3) and engagement (from item Y4) under a strict operationalization (top response category) and a soft one (half-credit for the intermediate category). The generating model is a two-dimensional graded response model [10] with a latent regression of both dimensions on the six covariates and language form, estimated by marginal maximum likelihood with Gauss–Hermite quadrature (13 nodes per dimension) [11] and automatic differentiation [12]; the correlation between dimensions is estimated freely. All settings, seeds, and code are in the released notebook wsdm_grm_market.ipynb, which regenerates the file byte-compatibly from the released data; the derived fields are conveniences, not primary data.

## Ethics

Both surveys were designed, fielded, and collected by Fledge.Love as voluntary, self-administered questionnaires distributed to its active users through an in-app prompt; participation carried no incentive, and declining carried no consequence. Two co-authors (D.Leshchikova and V. Klimov) contributed to the surveys' design and fielding in their capacity as platform employees. Anonymization, analysis, and the decision to publish were carried out independently by the academic authors (V. Kuskova and D. Zaytsev). Instrument B collected no direct identifiers; the contact field optionally collected by Instrument A was removed without being read. The de-identified data were transferred to the academic authors under a data-sharing agreement, and public release was authorized by the platform. The analysis and release protocol was reviewed by the University of Notre Dame Institutional Review Board and determined not to constitute human subjects research, as secondary analysis of de-identified data (protocol 26-08-10287). Privacy protection in the released files is provided by the anonymization pipeline described above, including the k ≥ 5 audit documented in the released log.

## Code availability

All code is released with the data under the MIT license [Data Citation 1]: build_release.py (the anonymization pipeline, which re-derives the analysis coding from the raw exports and asserts row-level equality before writing any file); three executable Jupyter notebooks: wsdm_grm_market.ipynb (measurement model, derived scores), wsdm_robustness.ipynb (differential item functioning, partial invariance, coding sensitivity, bootstrap), wsdm_validation.ipynb (descriptive paired analyses, latent class diagnostics, cross-validated predictive checks), and a table generator. The notebooks install their Python dependencies (autograd, scikit-learn, pandas, matplotlib) [12-14] at runtime and set fixed seeds at every stochastic step; the canonical outputs in results/ were produced by runs of these notebooks on the released data, and an independent re-execution reproduced cross-validation metrics to five decimal places. Each notebook ends with a verification cell that compares its outputs against results/ and reports agreement.

## Data Records

All records are archived together at Zenodo under DOI 10.5281/zenodo.21971273 [Data Citation 1], released under CC BY 4.0 (data) and MIT (code), with SHA-256 checksums for every file. The deposit contains three data files, an anonymization audit log, a bilingual codebook, the processing and analysis code, and the canonical outputs that the released code produces from the released data.

The file *instrument_b_agents.csv* holds the agent-receptivity survey: 2,617 records × 28 columns (one row per respondent), carrying the language form, ISO submission week, each item's verbatim response and analysis code, coded covariates, the two auxiliary categorical items, and demographic fields (Table 5). derived_scores.csv holds model-derived quantities for the 2,499 records complete on all items and covariates, linked by resp_id (Table 6). The file *instrument_a_genfeatures.csv* holds the generative-feature survey: 2,894 records × 14 columns from an unlinked sample (Table 7). Sample composition after recoding appears in Table 8; Table 9 accounts for every record from collection to the analysis set.

**Table 5. Column dictionary: instrument_b_agents.csv (2,617 records).**

| Column | Type | Values / range | Missing |
|---|---|---|---|
| resp_id | string | opaque identifier (B#####) | 0 |
| lang | string | ru / en | 0 |
| iso_week | string | ISO 8601 week, e.g. 2025-W47 | 0 |
| Y1_verbatim | string | 8 categories | 30 |
| Y1 | number | 0–2 | 30 |
| Y2_verbatim | string | 6 categories | 57 |
| Y2 | number | 0–2 | 57 |
| Y3_verbatim | string | 6 categories | 41 |
| Y3 | number | 0–2 | 41 |
| Y4_verbatim | string | 8 categories | 39 |
| Y4 | number | 0–2 | 39 |
| Y5_verbatim | string | 8 categories | 36 |
| Y5 | number | 0–3 | 36 |
| Y6_verbatim | string | 7 categories | 45 |
| Y6 | number | 0–2 | 45 |
| Y7_raw_1to10 | number | 1–10 | 40 |
| Y7 | number | 0–4 | 40 |
| female | number | 0–1 | 35 |
| gender | string | verbatim label, language-form specific | 35 |
| age | number | 0–3 | 30 |
| tenure | number | 0–2 | 29 |
| match_enough | number | 0–3 | 34 |
| conv | number | 0–3 | 33 |
| neg_emo | number | 0–2 | 40 |
| decay_reason | string | 6 categories per language form (verbatim, original language) | 44 |
| coping | string | 4 categories per language form (verbatim, original language) | 42 |
| lang_en | number | 0–1 | 0 |
| age_band | string | 18-24 / 25-34 / 35-44 / 45+ / 35+ (EN-form women, merged) | 30 |

**Table 6. Column dictionary: derived_scores.csv (2,499 records).**

| Column | Type | Values / range | Missing |
|---|---|---|---|
| resp_id | string | link key to instrument_b_agents | 0 |
| theta_send | number | -2.04214–2.25839 | 0 |
| theta_recv | number | -1.98544–2.17286 | 0 |
| p_deploy_strict | number | 3.77608e-05–0.999616 | 0 |
| p_deploy_soft | number | 0.000491661–0.9998 | 0 |
| p_engage_strict | number | 7.49005e-05–0.901953 | 0 |
| p_engage_soft | number | 0.000783243–0.948266 | 0 |

**Table 7. Column dictionary: instrument_a_genfeatures.csv (2,894 records).**

| Column | Type | Values / range | Missing |
|---|---|---|---|
| resp_id | string | opaque identifier (B#####) | 0 |
| iso_week | string | ISO 8601 week, e.g. 2025-W47 | 0 |
| idea1_profile | number | 0–2 | 143 |
| idea1_profile_nonresponse | number | 0–1 | 0 |
| idea2_tips | number | 0–2 | 236 |
| idea2_tips_nonresponse | number | 0–1 | 0 |
| idea3_couple | number | 0–2 | 107 |
| idea3_couple_nonresponse | number | 0–1 | 0 |
| best_of_three | string | 4 categories | 0 |
| interest_increase | number | 0–3 | 0 |
| gender | string | verbatim label, language-form specific | 0 |
| age_band | string | 18-24 / 25-34 / 35-44 / 45+ / 35+ (EN-form women, merged) | 0 |

| usage_freq | number | 0–2 | | 0 |
|---|---|---|---|---|
| ai_usage | number | 0–3 | | 0 |

**Table 8. Sample composition of Instrument B by language form, gender, and released age band (records with non-missing gender and age). No cell falls below k = 5.**

| Form | Gender | Age band | n |
|---|---|---|---|
| EN | Female (EN form) | 18-24 | 8 |
| EN | Female (EN form) | 25-34 | 16 |
| EN | Female (EN form) | 35+ | 12 |
| EN | Male (EN form) | 18-24 | 34 |
| EN | Male (EN form) | 25-34 | 96 |
| EN | Male (EN form) | 35-44 | 47 |
| EN | Male (EN form) | 45+ | 15 |
| RU | Female (RU form) | 18-24 | 191 |
| RU | Female (RU form) | 25-34 | 295 |
| RU | Female (RU form) | 35-44 | 261 |
| RU | Female (RU form) | 45+ | 129 |
| RU | Male (RU form) | 18-24 | 306 |
| RU | Male (RU form) | 25-34 | 644 |
| RU | Male (RU form) | 35-44 | 440 |
| RU | Male (RU form) | 45+ | 87 |

**Table 9. Record accounting, Instrument B.**

| Set | n |
|---|---|
| Instrument B records | 2617 |
| … complete on all seven items | 2517 |
| … complete on items and covariates (analysis set = derived scores) | 2499 |

The 118 records outside the analysis set remain in the released file with their observed values; missingness is confined to the columns shown in Table 5.

# Technical Validation

All quantities in this section are computed by the released notebooks from the released files; the canonical values live in results/ and regenerate on execution [Data Citation 1].

## Response quality and completeness

Of 2,617 Instrument B records, 2,517 are complete on all seven items and 2,499 on items and covariates jointly (Table 9); per-column missingness never exceeds 57 records (Table 5). Every response category of every item is used (Table 5 ranges), with no degenerate categories in either language form. Instrument A free-text nonresponse affects 143, 236, and 107 ratings on the three concepts respectively and is flagged per record.

## Measurement structure

The seven-item battery was designed to measure receptivity in two respondent roles, and the released data support that structure: a two-dimensional graded response model[10] (principal-role items Y1–Y3, Y7;

counterpart-role items Y4–Y6) fits better than a one-dimensional model by ΔBIC = 51.8 (likelihood-ratio statistic 114.5 on 8 df), with the two dimensions correlating 0.92 (bootstrap 95% CI [0.90, 0.94]). Item discriminations range from 1.72 (the 10-point value rating) to 4.20, indicating all items are strongly related to their assigned dimension; full item parameters are released (results/item_parameters_2d.csv).

## Cross-language comparability

Because the two forms are translations, we document where they function equivalently[15]. Item-level differential-functioning tests (each item's parameters freed for the English form against the joint model) flag two counterpart-role items, Y4 (LRT 15.4, 3 df, $p = .001$) and Y5 (LRT 41.3, 4 df, $p < 10^{-7}$), and none of the principal-role items. By gender, only Y5 is flagged (LRT 16.6). A partial-invariance re-fit freeing the flagged items leaves the remaining parameters essentially unchanged (released in results/partial_invariance.csv). Users comparing scores across forms should treat Y4 and Y5 as non-invariant and either free them, as the released notebook does, or restrict cross-form comparisons to the invariant items; the English form is also small (232 records), so form-level contrasts carry wide uncertainty.

## Parameter stability and precision

A 200-replicate nonparametric bootstrap (reduced-quadrature replicates recentered on full-precision estimates; seed scheme documented in the notebook) yields the confidence intervals released in results/bootstrap_ci.csv; no reported parameter interval is degenerate. Two items admit defensible alternative category orderings. Refitting under each (results/sensitivity.csv) moves the separation between the two focal item thresholds within a 0.02 range (0.70–0.72), and the dimension correlation within 0.001. A latent class analysis over the seven items selects four classes by BIC (27,632, against 28,031 at K = 3 and 27,633 at K = 5) with relative entropy 0.83; all 20 random restarts recover the identical solution, and the class structure persists under K = 5 (98.9% of the largest distinctive class maps to a single K = 5 class of the same profile). These diagnostics are evidence that the population structure recoverable from the data is stable under estimation choices.

## Predictive coherence

As an internal-consistency check, we ask whether a held-out item is predictable from the remainder of the battery. Under five-fold respondent-level cross-validation, with item Y4 excluded entirely from model fitting, scores are estimated from the six remaining items and covariates, with a logistic link calibrated on the training folds. Out-of-fold prediction of Y4 achieves AUCs of 0.89 for any endorsement and 0.88 for full endorsement, with Brier scores of 0.132 and 0.079, respectively, compared with base-rate references of 0.240 and 0.107. Simple benchmarks on identical folds (a covariates-only classifier: AUC 0.60; a raw two-item sum score [16]: 0.85; a supervised logistic on the six items: 0.89) bracket the latent score, indicating the battery's items share a coherent construct rather than idiosyncratic variance (presented in files *results/routing_validation.csv*, *results/routing_baselines.csv*).

### Reproducibility verification

Every file in results/ was produced by executing the released notebooks on the released data. An independent local re-execution of the cross-validation reproduced the published AUCs to five decimal places, and each notebook's final cell re-derives its outputs and compares them to results/ with explicit tolerances. SHA-256 checksums cover every file in the deposit.

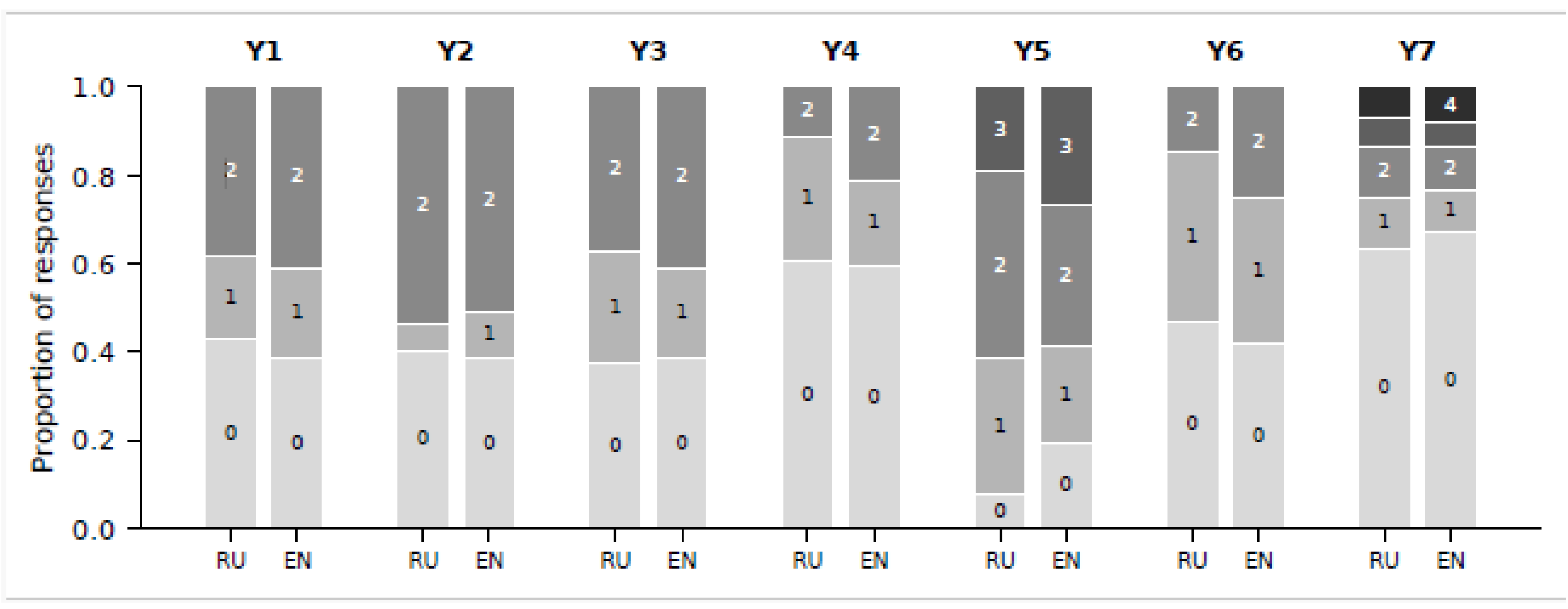

***Figure 2: Response distributions for the seven receptivity items by language form (category shares; codes as in Tables 1–2). Every category of every item is used in both forms.***

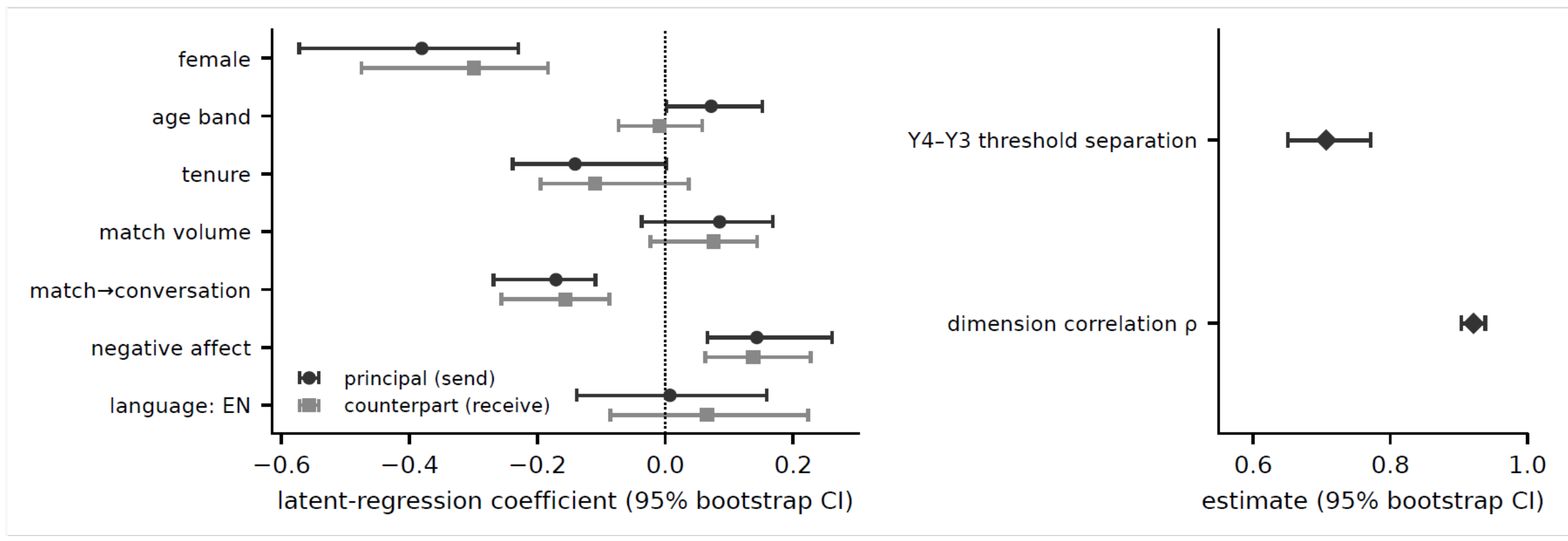

***Figure 3 (near here). Precision of released model parameters: latent-regression coefficients for both dimensions (left) and the two summary parameters (right), with 200-replicate bootstrap 95% intervals as released in results/bootstrap_ci.csv.***

## Usage Notes

**Getting started**. The three CSVs load directly (UTF-8; comma-separated). Link *derived_scores.csv* to *instrument_b_agents.csv* on resp_id. the Instrument A sample is not linkable to Instrument B by design. To regenerate every derived quantity, run the notebooks in order: *wsdm_grm_market.ipynb*, *wsdm_robustness.ipynb, wsdm_validation.ipynb*, pointing their BASE variable at the directory containing the data files. Each installs its own dependencies and ends with a verification cell.

**Comparability guidance**. Scores are comparable across the two language forms only under the partial-invariance configuration (items Y4 and Y5 freed); the released notebook implements it, and results/dif_results.csv documents the tests. The English subsample is small (n = 232), and for English-form women the two oldest age bands are merged ("35+") with the numeric age code recoded accordingly for three records; analyses stratifying English-form respondents by fine age bands are not supported by the release.

**Limitations**. All receptivity measures are stated preferences collected by survey; the release contains no behavioral logs, no message content, and no outcomes of actual agent interactions. The sample is one platform's active users reached by in-app prompt; engaged users are overrepresented (80.1% of Instrument A respondents report opening the app several times a week), and the two samples are cross-sectional and unlinked. Free-text responses were collected but are not released.

**Reuse directions.** (1) Human–AI communication: the battery separates principal-role from counterpart-role receptivity within respondents, supporting measurement work on delegated communication. (2) Recommender systems: the per-user engagement propensities in derived_scores.csv can serve as targets or benchmarks for agentic and reciprocal recommendation models [9]. (3) Cross-cultural technology acceptance: a bilingual instrument with documented item-level functioning differences supports translation and invariance methodology [15]. (4) Teaching: the dataset is small, complete, and ships with a full graded-response, invariance, and latent-class pipeline with published outputs, making it a self-checking exercise set for psychometrics courses.

**Licensing and citation**. Data are CC BY 4.0; code is MIT. Please cite this Data Descriptor and the deposit [Data Citation 1].

**Author Contributions.** D.Leshchikova, V. Klimov: designed and fielded the instruments. V.Kuskova, D. Zaytsev: designed the release, performed the anonymization and all analyses, wrote the code and the manuscript.

**Competing Interests.** D.Leshchikova and V. Klimov are employees of Fleamily, Inc., which operates Fledge.Love. The surveys were designed and fielded by the platform. The anonymization, analyses, validation, and the decision to publish were carried out independently by the academic authors (V.Kuskova, D. Zaytsev). The authors declare no other competing interests.

**Acknowledgements.** We are grateful to Fledge.Love for releasing the data for public use.

**Funding.** None.

**Data Citations.** [Data Citation 1] Zenodo https://doi.org/10.5281/zenodo.21971273 (2026).